\documentclass[trackchanges,twocolumn]{aastex}

\let\tablenum\relax

\usepackage{natbib,epsfig,siunitx,url,graphicx,amsmath,footnote,float,xcolor,cases}
\begin{document}

\submitted{Accepted for publication in Icarus}

\title{Born extra-eccentric: A broad spectrum of primordial configurations of the gas giants that match their present-day orbits}

\author{Matthew S. Clement\altaffilmark{1}, Rogerio Deienno\altaffilmark{2}, Nathan A. Kaib\altaffilmark{3}, Andr\'{e} Izidoro\altaffilmark{4}, Sean N. Raymond\altaffilmark{5} \& John E. Chambers\altaffilmark{1}}

\altaffiltext{1}{Earth and Planets Laboratory, Carnegie Institution for Science, 5241 Broad Branch Road, NW, Washington, DC 20015, USA}
\altaffiltext{5}{Laboratoire d'Astrophysique de Bordeaux, Univ. Bordeaux, CNRS, B18N, all{\'e} Geoffroy Saint-Hilaire, 33615 Pessac, France}
\altaffiltext{3}{HL Dodge Department of Physics Astronomy, University of Oklahoma, Norman, OK 73019, USA}
\altaffiltext{2}{Southwest Research Institute, 1050 Walnut St. Suite 300, Boulder, CO 80302, USA}
\altaffiltext{4}{Department of Earth, Environmental and Planetary Sciences, MS 126, Rice University, Houston, TX 77005, USA}
\altaffiltext{*}{corresponding author email: mclement@carnegiescience.edu}

\begin{abstract}

In a recent paper we proposed that the giant planets' primordial orbits may have been eccentric ($e_{J} \sim e_{S} \sim$ 0.05), and used a suite of dynamical simulations to show outcomes of the giant planet instability that are consistent with their present-day orbits.  In this follow-up investigation, we present more comprehensive simulations incorporating superior particle resolution, longer integration times, and eliminating our prior means of artificially forcing instabilities to occur at specified times by shifting a planets' position in its orbit.  While we find that the residual phase of planetary migration only minimally alters the the planets' ultimate eccentricities, our work uncovers several intriguing outcomes in realizations where Jupiter and Saturn are born with extremely large eccentricities ($e_{J}\simeq$ 0.10; $e_{S}\simeq$ 0.25).  In successful simulations, the planets' orbits damp through interactions with the planetesimal disk prior to the instability, thus loosely replicating the initial conditions considered in our previous work.  Our results therefore suggest an even wider range of plausible evolutionary pathways are capable of replicating Jupiter and Saturn's modern orbital architecture.

\end{abstract}

\section{Introduction}

Dynamical interactions between the young giant planets played a crucial role in molding our solar system's global properties.  As seems to be common for giant exoplanets \citep[e.g.:][]{marois08,rivera10,hr8788,bae19}, interactions with the primordial nebular gas likely conspired to corral the solar system's giants into a compact chain of resonant orbits \citep{masset01,morby07}.  The \textit{Nice Model} \citep{Tsi05,gomes05,mor05} describes how the cataclysmic destruction of this conglomeration of harmonized orbits \citep{morby09,nesvorny11,nesvorny12} successfully generates many peculiar qualities of the solar system \citep[see][for a recent review]{nesvorny18_rev}.  While numerous contemporary studies have found various observed structures in the solar system to be \textit{consistent} with such an event, the lack of a compelling alternative explanation for irregular satellite captures around all four giant planets \citep{nesvorny14a,nesvorny14b} and certain properties of the modern asteroid belt \citep{walshmorb11,minton11,clement20_mnras} arguably \textit{necessitate} the occurrence of an instability in the solar system's past.

Constraining the instability's precise timing within the larger sequence of events transpiring during the solar system's formative epochs (i.e.: the amount of time the resonant chain survived prior to destruction) has sparked a sizable literary output over the past several years.  While classic studies \citep[e.g.:][]{gomes05,levison11} argued that the instability provoked the late heavy bombardment\footnote{A perceived spike in the Moon's cratering history $\sim$650 Myr after gas dispersal; the existence of which has been called into question in recent years \citep{tera74,zellner17}.}, contemporary work tends to favor the event's transpiration within the first $\sim$100 Myr after the solar system's birth \citep{morb18,quarles19}.  Indeed, certain distinctive features including binary trojan satellites of Jupiter \citep{nesvorny18}, asteroid families with inferred ages $\gtrsim$4.5 Gyr \citep{delbo17,delbo19}, and the terrestrial planets' dynamical excitation, masses, and compositions broadly suggest an earlier version of the Nice Model \citep{clement18,clement18_frag,clement18_ab,deienno18,mojzsis19,brasser20,nesvorny21_tp,woo21_tp}.  In light of the deduced significance of the instability's specific timing, dynamical models often incorporate artificial instability triggers to ensure the event initiates at the appropriate time \citep[e.g.:][]{clement18,clement21_instb} and minimize the computational cost of the calculation \citep{nesvorny11,nesvorny12}.  

While a tenuous consensus in favor of an early instability has developed in the past several years, the connection between disk model predictions of the giant planets' emergent orbits and their present-day configuration remains somewhat dubious.  Given an appropriate combination of prescribed disk parameters, modern models \citep[e.g.:][]{pierens08,zhang10,dangelo12,pierens14} generally find that Jupiter and Saturn can be captured in either a 2:1 or 3:2 mean motion resonance (MMR), and experience either inward or outward migration \citep[in some cases both:][]{pierens11}.  While two-phase migration is advantageous for limiting the mass of Mars and the asteroid belt \citep{walsh11,jacobson14,brasser16}, other explanations for these qualities do not require migration \citep[e.g.:][]{levison15,izidoro15,clement18,deienno18}.

Understanding how the ice giants' formation and early migration fits into this story remains an outstanding puzzle.  This gap is unfortunate given that the mutual interactions between Jupiter, Saturn, and the first adjacent ice giant considerably influence the planets' ultimate eccentricities and semi-major axes \citep{nesvorny12,clement21_instb}, and Neptune's pre-instability migration equally modifies the young Kuiper Belt \citep{nesvorny15a,nesvorny15b,kaib16,gomes18,volk19}.  Of particular relevance to the Nice Model discussion, a delayed phase of giant impacts on Uranus and Neptune has been proposed as a mechanism to self-consistently replicate the ice worlds' obliquities \citep{izidoro15_ig}.  Recent work demonstrated that such a scenario often yields unstable resonant chains around the time of gas dispersal; thus potentially providing a trigger for an early instability \citep{ribeiro20}.   

In this paper we focus our attention on Jupiter and Saturn's acquisition of their particular modern secular architecture \citep{morby09}.  Specifically, our present work is a direct follow-up to our recent article on the topic \citep[][henceforth referred to as Paper I]{clement21_instb} where we argued that the primordial 2:1 Jupiter-Saturn resonance is advantageous in terms of its ability to consistently replicate the gas giants' eccentricities.  We highly encourage the reader to consult the background sections and appendices in Paper I for a more detailed synopsis of the present state of the field, and a more through summary of the secular theory of dynamical evolution in the solar system \citep{poincare1892}.  The Lagrange-Laplace solution describes how the planets' mutual perturbations within the N-body problem facilitate the eccentric and nodal precession terms in their orbits:
\begin{equation}
    \begin{array}{l}
    e_{i}\cos{\varpi_{i}} = \sum_{j}^{8}M_{ij}\cos{(g_{j}t + \beta_{j})} \\
    e_{i}\sin{\varpi_{i}} = \sum_{j}^{8}M_{ij}\sin{(g_{j}t + \beta_{j})} \\
	\label{eqn:sec}
	\end{array}
\end{equation}
As Jupiter and Saturn are the most massive planets, their respective dominant eigenfrequencies $g_{5}$ and $g_{6}$ are key drivers of solar system's dynamical evolution as a whole.  The frequencies themselves are largely determined by the planets' radial configuration, and thus are reasonably reproduced in any instability model that yields the correct giant planet semi-major axes \citep{morby09,nesvorny12}.  Conversely, the magnitudes of each frequency in the various planets' eccentricities ($M_{ij}$) were acquired via gravitational encounters (i.e.: encounters occurring at the time of the instability).  Thus, it is essential for instability models to consistently reproduce these dominant amplitudes; namely those of the Jupiter-Saturn system ($M_{55}$ and $M_{56}$ for Jupiter's eccentricity and $M_{65}$ and $M_{66}$ for Saturn's).  High values of $M_{55}$ can be achieved in simulations where the gas giants are placed in an initial 3:2 MMR if Jupiter experiences a close encounter with another planet \citep{morby09,nesvorny11}. However, it is noticeably difficult to simultaneously match Saturn's semi-major axis (typically $a_{S}$ is exceeded), eccentricity ($M_{65}$ and $M_{66}$ are often over-excited), and forcing on Jupiter (usually final $M_{56}$ values are too large) in such a scenario.  Consequently, the solar system outcome lies at the extreme limit of possible numerically generated Jupiter-Saturn orbital spacings and $M_{55}$ values in the 3:2 version of the Nice Model.  

In Paper I we found that the actual Jupiter-Saturn system represents a more typical numerical result when the giants originate in a mutual 2:1 resonance.  In particular, the 2:1 is advantageous since the planets attain inflated eccentricities during the nebular disk phase as they carve out larger mutual gaps in the gas \citep[leading to weaker tidal damping on their eccentricities:][]{pierens14}.  Indeed, hydrodynamical investigations of both the solar system \citep[e.g.:][]{zhang10,pierens11} and giant exoplanet systems \citep[e.g.:][]{kley04,bae19} find highly eccentric 2:1 outcomes given particular combinations of disk parameters.  The current version of the Nice Model invokes the existence of an additional one or two ice giants to maximize the chance of finishing with the correct number of planets \citep{nesvorny11} and minimize the time in which powerful resonances inhabit certain regions of the inner solar system \citep{bras09,minton11,walshmorb11}.  In Paper I we demonstrated that Uranus and Neptune's final orbits can be fine-tuned to more closely resemble the real ones by adjusting the total mass of the primordial Kuiper Belt, and that of the ejected ice giant \citep[essentially free parameters in investigations of the instability:][]{nesvorny12}.  However, in order to perform a wide parameter space sweep of the plausible 2:1 resonant chains, the simulations in Paper I were limited in resolution and duration.  In this sequel manuscript we present lengthier and higher-resolution integrations of select sets of initial conditions found to be successful in Paper I.  Additionally, we perform a batch of computations without artificially triggering instabilities (a method we employed to originate each simulation from the same $e_{J}$/$e_{S}$ combination in Paper I).  Notably, this new set of self-triggered instabilities consider more extreme initial eccentricities for the planets that are consistent with the results of disk models \citep{pierens14}.  

Our present investigation addresses two important problems left unresolved in Paper I.  First, the effects of residual migration beyond 20 Myr (i.e.: migration driven by interactions with remaining Kuiper-belt planetesimals, some of which are likely still present) on Saturn's position relative to its 5:2 MMR with Jupiter and the eccentric amplitudes $M_{ij}$ remain unclear.  As the chief advantage of the primordial 2:1 Jupiter-Saturn resonance is its tendency to more consistently reproduce these values in numerical simulations, it is important to quantitatively understand the degree to which they might be negatively altered via residual migration.  Second, the connection between our simulated initial conditions in Paper I and disk model predictions \citep{pierens14} of Jupiter and Saturn's orbital configuration within the primordial 2:1 MMR remains somewhat vague.  In particular, the planets attain significantly higher initial eccentricities while engulfed in the nebular gas than the values of $e_{J}$ and $e_{S}$ tested in Paper I.  The ability of these 2:1 Jupiter-Saturn chains to match constraints related to their present-day spacing and partitioning of eccentric secular modes is directly related to this primordial excitation.  Therefore, it is important to concretely bridge the gap between \citet{pierens14} and Paper I with simulations designed to test the viability of extremely inflated initial gas giant eccentricities.

\section{Methods}

\subsection{Numerical Simulations}
\label{sect:cpu}

We direct the reader to the methods sections of Paper I for a more comprehensive description of our computational pipeline.  In short, each of our simulations leverage the $Mercury6$ numerical integration package \citep{chambers99}, employ a 50.0 day time-step, remove objects that make perihelion passages less than 0.1 au, and consider particles ejected at heliocentric distances of 1,000 au.  Resonant chains are generated with fictitious forces implemented to mimic gas disk interactions via forced migration and eccentric damping terms \citep[e.g.:][]{lee02}.  Once in resonance, we excite the eccentricities of the planets by either reducing the magnitude of the eccentric damping force, or reversing its sign.  In all cases presented here we integrate the giant planet configurations in the absence of external forces or additional particles (i.e.: the primordial Kuiper Belt Objects; KBOs) for 1 Myr to ensure a degree of stability and check for resonant libration prior to initiating our production runs.  When distributing KBOs exterior to our resonant giant planets, we assign the particles identical masses, semi-major axes such that the surface density profile falls off as $r^{-1}$, eccentricities and inclination drawn from near-circular Rayleigh distributions ($\sigma_{e}=$ 0.01, $\sigma_{i}=$ 1$^{\circ}$), and select the remaining orbital elements by randomly sampling uniform distributions of angles.  In all cases the disk's inner edge is radially offset from the most distant planet by 1.5 au, and the outer edge is at 30.0 au \citep[see Paper I and:][]{gomes04,ribeiro20}.  As in Paper I, instabilities are triggered via an abrupt shift in the mean anomaly of the innermost ice giant if a system has not destabilized after 100.0 Kyr of simulation time (with the exception of the simulations described in section \ref{sect:q3} where an artificial trigger is not used).

\subsection{Fully resolved residual migration phase}
\label{sect:q1}

Table \ref{table:ics} summarizes the initial conditions and total number of simulations analyzed for our present study, in addition to three sets of computations from Paper I utilized in this manuscript for the purposes of comparison.  In Paper I we integrated each system for 20.0 Myr after the onset of the instability to capture some degree of the residual phase of migration \citep{clement20_mnras}.  However, depending on the amount of mass remaining in the post-instability Kuiper Belt, appreciable migration can continue for $\sim$100 Myr \citep{nesvorny12,brasser_lee_15}.  While computationally intensive, it is important to study this more complete epoch of post-instability evolution as the planets' eccentricities tend to damp via dynamical friction throughout the phase.  Moreover, if residual migration is indeed significant, the planets' final semi-major axes can change substantially.  If this is the case, certain simulations deemed ``successful'' or ``unsuccessful'' in Paper I might evolve towards the opposite designation.  For this reason, we perform 400 simulations considering a successful set\footnote{Note that this particular set of initial conditions is specifically selected because it produced the largest sample of 4 planet systems with $P_{S}/P_{J}<$ 2.5 in Paper I.  This allows us to maximize our sample of solar system-like post-instability configurations for integration through the residual migration phase.} of initial conditions from Paper I (a six planet, 2:1,4:3,4:3,3:2,3:2, $e_{J}=e_{S}=$ 0.05 chain) that fully resolve the residual migration phase for 100 Myr following the onset of the instability.  Half of these simulations set the total mass of the primordial Kuiper Belt to 20.0 $M_{\oplus}$ (the nominal value determined in Paper I), and half investigate a disk mass of 40.0 $M_{\oplus}$ (a more extreme case in terms of the presumed effects on residual migration).  As we determined that an ejected ice giant mass of $M_{IG}=$ 6.0 $M_{\oplus}$ markedly improved simulation results for six planet configurations in Paper I, we utilize this selection of $M_{IG}$ for all six planet chains presented in this manuscript.  Thus, these simulations consider a chain of ice giants with successive masses of 6.0, 6.0, 16.0 and 16.0 $M_{\oplus}$.

\begin{table*}
	\centering
	\begin{tabular}{c c c c c c c c c}
	\hline
	Name & Resonant Chain & $e_{J,o}$ & $e_{S,o}$ & $M_{KB}$ ($M_{\oplus}$) & $M_{IG}$ ($M_{\oplus}$) & $t_{int}$ (Myr) & $N_{KBO}$ & $N_{sim}$ \\
	\hline
	\textbf{Comparison from Paper I} &&&&&&\\
	\hline
	Control & 2:1,4:3,4:3,3:2,3:2 & 0.05 & 0.05 & 20.0 & 8.0 & 20.0 & 1,000 & 183 \\
	Low $M_{KB}$ & 2:1,4:3,4:3,3:2,3:2 & 0.05 & 0.025 & 20.0 & 6.0 & 20.0 & 1,000 & 183 \\
	High $M_{KB}$ & 2:1,4:3,4:3,3:2,3:2 & 0.05 & 0.025 & 40.0 & 8.0 & 20.0 & 1,000 & 186 \\
    \hline
    \textbf{This work} &&&&&&\\
    \hline
	100 Myr Low $M_{KB}$ & 2:1,4:3,4:3,3:2,3:2 & 0.05 & 0.05 & 20.0 & 6.0 & 100.0 & 1,000 & 188 \\
	100 Myr High $M_{KB}$ & 2:1,4:3,4:3,3:2,3:2 & 0.05 & 0.05 & 40.0 & 6.0 & 100.0 & 1,000 & 197 \\
	High-res & 2:1,4:3,4:3,3:2,3:2 & 0.05 & 0.05 & 20.0 & 6.0 & 20.0 & 5,000 & 181 \\
	5GP High-$e$ & 2:1,2:1,3:2,3:2 & 0.09 & 0.22 & 20.0 & 8.0 & 20.0 & 1,000 & 92 \\
	6GP High-$e$ & 2:1,4:3,4:3,3:2,3:2 & 0.08 & 0.23 & 20.0 & 8.0 & 20.0 & 1,000 & 90 \\
	\hline
	\end{tabular}
	\caption{Summary of initial conditions for our present manuscript, as well as those for simulations from Paper I used for comparison.  The columns are as follows: (1) the name of the simulation set, (2) the resonant chain tested, (3-4) the initial eccentricities of Jupiter and Saturn, (5) the total mass of the primordial external planetesimal disk, (6) the mass of the innermost ice giant(s), (7) the total integration time after the onset of the instability, (8) the number of particles used to represent the primordial planetesimal disk, and (9) the number of simulations in each batch producing an instability and retaining both Jupiter and Saturn (in almost all cases the reason the total of runs is not equal to 100 or 200 is that an instability did not occur).}
	\label{table:ics}
\end{table*}

\subsection{Improved planetesimal disk particle resolution}
\label{sect:q2}

An additional simplification made in Paper I was to represent the primordial Kuiper Belt with 1,000 equal-mass objects (in the majority of our simulations each KBO's mass was approximately ten times that of Pluto).  Thus, the ability of our simulations to properly resolve the external disk's gravitational perturbations on the resonant giant planets was inadequate in terms of artificially boosting the typical encounter strengths.  While the exact make-up of the primeval belt remains a subject of ongoing debate \citep[see][for a recent review]{morby20_kb_rev}, gravitational interactions with Pluto-mass bodies play a crucial role in sculpting the specific structure of the Kuiper Belt by facilitating a migration history of Neptune that is grainy rather than smooth 
\citep[characterized by stochastic jumps in semi-major axis:][]{nesvorny16,kaib16}.  In this manner, \citet{nesvorny16} concluded that the nominal young Kuiper Belt contained $\sim$4,000 Pluto-mass bodies.  Moreover, such a primordial size frequency distribution is potentially consistent with the formation \citep{canup05} and inferred cratering history of the Pluto-Charon system \citep{kenyon20}, as well as the the genesis of their mutual satellite system \citep{bromley20}.  While previous studies have found the effects of varying the number of primordial KBOs on the instability evolution of Jupiter and Saturn to be minimal beyond $N_{KBO} \simeq$ 1,000 \citep{levison11,nesvorny12,quarles19}, for the purposes of comparison we repeat the simulations described in the previous subsection with 5,000 disk members (note, however, that our simulations do not consider the effects of KBO self-gravity).  As our 100 Myr simulations indicate that residual migration beyond $\sim$20 Myr minimally affects the final statistical distributions of simulation results (discussed in section \ref{sect:res_mig}), we utilize a 20 Myr post-instability integration time for the remainder of our investigation.

\subsection{High-eccentricity runs}
\label{sect:q3}

As in Paper I, the simulation sets described in sections \ref{sect:q1} and \ref{sect:q2} apply an artificial instability trigger (an abrupt shift in the mean anomaly of the innermost ice giant) to ensure each instability ensues from a determined combination of $e_{J}$ and $e_{S}$.  While this methodology is advantageous when attempting to ``reverse engineer'' the planets' modern partitioning of eccentric secular modes, the connection between our results and gas disk models is consequentially somewhat vague.  Specifically, the best performing sets of initial conditions tested in Paper I originated with $e_{J}=e_{S}=$ 0.05 (five planet chains) and $e_{J}=$ 0.05, $e_{S}=$ 0.025 (six planet chains).  

In the absence of an artificial instability trigger, the giant planets' orbits tend to damp towards near-zero eccentricity as they smoothly migrate prior to the instability's inception \citep{nesvorny12}.   Thus, the viability of higher \textit{initial} values of $e_{J}$ and $e_{S}$ is still unclear.  Investigating such eccentric chains of orbits is particularly compelling as hydrodynamic models of the gas giants' evolution within the 2:1 MMR in the nebular disk phase find initial eccentricities of order $e_{J} \simeq$ 0.10 and $e_{S} \simeq$ 0.25.  Similarly, simulations investigating PDS-70 (perhaps the quintessential example of two young giant planets potentially evolving in a proto-planetary disk within the 2:1 resonance) found that eccentricities of $\sim$0.025-0.10 for the inner planet and $\sim$0.10-0.35 for the outer planet best replicating the observed disk structure \citep{bae19}.  Though the overall differences between our results for various $e_{J}/e_{S}$ combinations in Paper I were minor, all of our presumed configurations were still mildly inconsistent with the structures generated in disk models.  

In order to investigate whether more eccentric initial configurations might damp towards more moderate values of $e_{J}$ and $e_{S}$ via smooth migration prior to the instability, we perform two sets of simulations without an instability trigger where the gas giants' eccentricities are akin to those found in \citet[][see table \ref{table:ics}]{pierens14}.  In general, we construct these chains such that the outermost ice giant's semi-major axis is similar to values found successful in Paper I \citep[see also:][]{batbro12,nesvorny12,deienno17}.  Therefore, we utilize a looser chains of resonances for our set investigating a five planet configurations (2:1,2:1,3:2,3:2), and leverage a more compact configuration in our 6GP, High-$e$ batch (2:1,4:3,4:3,3:2,3:2).  It is worth noting that the innermost ice giant in these more compact, eccentric six planet chains begin each simulation on a crossing orbit with Saturn that is phase-protected from collisional trajectories while the planets remain in resonance.  While it is unclear whether such a chain of resonant planets might have emerged from the nebular gas in such an overlapping configuration, these simulations present an interesting and more exotic comparison to the more conventional parameters explored in Paper I.

\subsection{Success Criteria}
\label{sect:success}

We leverage the same success criteria as in Paper I, which were largely motivated by the four constraints developed in \citet{nesvorny12}.  Criterion \textbf{A} requires that a system finish with exactly four planets ($N_{GP}=$ 4).  Provided criterion \textbf{A} is satisfied, we assess the broad radial structure of the resulting outer solar system with criterion \textbf{B}; which stipulates that each successive planets' semi-major axis complete the simulation within 20$\%$ of the real value.  Similarly, criterion \textbf{C} requires each of the four eccentric magnitudes of the Jupiter-Saturn system ($M_{ij}$) finish within 50$\%$ of the real value, and the integration conclude in the $M_{55}>M_{56}$ regime.  Finally, criterion \textbf{D} separates simulations where Jupiter and Saturn remain inside of the 5:2 MMR \citep[$P_{S}/P_{J}<$ 2.5, see:][]{clement20_mnras} from those that do not.  

In Paper I we discussed the efficacy of assessing simulation success with a small number of broad constraints that may or may not have mutual exclusivities.  As the instability is highly stochastic, a sufficiently large batch of numerical simulations assuming near-identical initial conditions is apt to yield a diverse spectrum of evolutionary outcomes.  While it is philosophically appealing to favor a suite of computations that delivers a sizable population of architectures akin to the modern giant planet configuration, it is equally possible that the solar system resulted from a low-probability chain of events (though we are not arguing for such a scenario here).  Moreover, a set of simulations yielding no outcomes that simultaneously satisfy all four success criteria might still be successful if the shortcoming is the result of small number statistics and an over-multiplication of constraints.  For these reasons, along with those outlined in section 2.4 of Paper I, we concentrate our assessment of our results on both the four constraints themselves, and the various mutual exclusivities that arise between them.

\section{Results}

Table \ref{table:results} lists each simulation's success when scrutinized against our four metrics established in the previous section.  In the subsequent sections, we summarize our major findings for each of the three main investigations and open-ended questions from Paper I described in sections \ref{sect:q1}, \ref{sect:q2} and \ref{sect:q3}.

\begin{table*}
	\centering
	\begin{tabular}{c c c c c c c c c c c}
	\hline
	Name & $e_{J,o}$ & $e_{S,o}$ & $M_{KB}$ ($M_{\oplus}$) & $t_{int}$ (Myr) & $N_{KBO}$ & \textbf{A} & \textbf{B} & \textbf{C} & \textbf{D} & \textbf{ALL}   \\
	\hline
	\textbf{Comparison from Paper I} &&&&&&\\
	\hline
	Control & 0.05 & 0.05 & 20.0 & 20.0 & 1,000 & 54 & 26 & 14 & 31 & 1 \\
	Low $M_{KB}$ & 0.05 & 0.025 & 20.0 & 20.0 & 1,000 & 60 & 37 & 10 & 57 & 2  \\
	High $M_{KB}$ & 0.05 & 0.025 & 40.0 & 20.0 & 1,000 & 57 & 24 & 11 & 23 & 0 \\
    \hline
    \textbf{This work} &&&&&&\\
    \hline
	100 Myr Low $M_{KB}$ & 0.05 & 0.05 & 20.0 & 100.0 & 1,000 & 64 & 27 & 13 & 30 & 2 \\
	100 Myr High $M_{KB}$ & 0.05 & 0.05 & 40.0 & 100.0 & 1,000 & 45 & 10 & 9 & 19 & 0 \\
	High-res & 0.05 & 0.05 & 20.0 & 20.0 & 5,000 & 46 & 21 & 11 & 32 & 1 \\
	5GP High-$e$ & 0.09 & 0.22 & 20.0 & 20.0 & 1,000 & 25 & 21 & 12 & 15 & 2 \\
	6GP High-$e$ & 0.08 & 0.23 & 20.0 & 20.0 & 1,000 & 38 & 23 & 10 & 15 & 2 \\
	\hline
	\end{tabular}
	\caption{Summary of results and key statistics for our various simulation sets.  The columns are as follows: (1) the name of the simulation set, (2-3) the initial eccentricities of Jupiter and Saturn, (4) the total mass of the primordial external planetesimal disk, (5) the total integration time after the onset of the instability, (6) the number of particles used to represent the primordial planetesimal disk, (7) the percentage of systems satisfying criterion \textbf{A} ($N_{GP}=$4), (8) criterion \textbf{B} (the planets' final semi-major axes within 20$\%$ of the real ones), (9) criterion \textbf{C} ($|\Delta M_{ij}/M_{ij,ss}|<$0.50 ($i,j=$5,6), $M_{55}>M_{56}$), (10) criterion \textbf{D} ($P_{S}/P_{J}<$2.5), and (11) the percentage of systems satisfying all four success criteria simultaneously.}
	\label{table:results}
\end{table*}

\subsection{Fully resolved residual migration phase}
\label{sect:res_mig}

Our 100 Myr simulations indicate that residual migration beyond $t_{inst}$ $+$ 20 Myr only minimally alters our final system architectures.  In general, the eccentric magnitudes of the Jupiter and Saturn system damp strongly in the 5-10 Myr interval of migration following Jupiter's jump.  Subsequent reduction in the amplitudes $M_{ij}$ is minimal, and limited to the $\sim$10$\%$ level in the majority of our simulations. Comparing the final statistics generated in our two batches of simulations that fully capture the planets' residual migration phase to those of the similar sets from Paper I (table \ref{table:results}: \textit{Low $M_{KB}$} for a close analog to our \textit{100 Myr Low $M_{KB}$} set and \textit{High $M_{KB}$} for a close analog to our \textit{100 Myr High $M_{KB}$} set), the most obvious discrepancy is the surprisingly low rates of success for criteria \textbf{A} ($N_{GP}$) and \textbf{B} (the planets' semi-major axes) in our extended simulations investigating $M_{KB}=$ 40.0 $M_{\oplus}$ disks.  However, this shortcoming is directly attributable to the fact that over half of these simulations (108 of 187 systems) undergo relatively weak instabilities and finish with $N_{GP}=$ 5.  We find similar trends in our comparison simulation batch from Paper I (\textit{High $M_{KB}$}: 65 of 186 simulations finishing with five total giant planets).  As we would not expect the different total integration times to affect the surviving number of planets positively, the inconsistency between these statistics is possibly due to the disparate ice giant masses (8.0 versus 6.0 $M_{\oplus}$).  

The tendency of higher values of $M_{KB}$ to boost the total number of surviving planets by supplying increased dynamical friction that tends to damp the excited ice giant orbits has been noted before \citep[e.g.:][]{nesvorny12}.  Thus, the low success rates for \textbf{A} and \textbf{B} only speak against the specific choice of $M_{KB}=$ 40 $M_{\oplus}$ for these particular chains.

We note substantial differences between the reference systems from Paper I and our new simulations in terms of the the cumulative number of systems satisfying criteria \textbf{B} and \textbf{D} (our constraints for the giant planets' semi-major axes and $P_{S}/P_{J}$, respectively).  This is an obvious consequence of residual migration transforming successful simulations to unsuccessful ones, and is particularly pronounced in our simulations testing $M_{KB}=$ 40 $M_{\oplus}$ (the expected result of higher post-instability disk masses).  Simply put, the planets' semi-major axes evolve too far past those of the real giant planets.  While the results for \textbf{B} drop from 37$\%$ to 27$\%$ (\textit{Low $M_{KB}$}) and 24$\%$ to 10$\%$ (\textit{High $M_{KB}$}) when post-instability migration is fully resolved, the difference in success rates for \textbf{D} are rather substantial (57$\%$ versus 30$\%$ and 19$\%$).  This is not particularly surprising given that criterion \textbf{D} is more difficult to match, and that it provides no tolerance for systems exceeding the solar system value.  As a reasonable fraction of systems (30$\%$) complete the residual migration interval inside of $P_{S}/P_{J}=$ 2.5 \citep{clement20_mnras} in simulations investigating our preferred initial planetesimal disk mass of 20.0 $M_{\oplus}$, we conclude that the viability of our scenario of Jupiter and Saturn's capture in the primordial 2:1 resonance is not strongly dependent on a particular duration of ultimate migration as the effects of $M_{KB}$ are far more significant \citep{nesvorny12}.  In addition to the solar system value of $P_{S}/P_{J}$ falling well within the spectrum of outcomes produced in our simulations, the fraction of systems satisfying all four criterion (2$\%$) is identical to that of our reference batches from Paper I.

\begin{figure*}
	\centering
	\includegraphics[width=.45\textwidth]{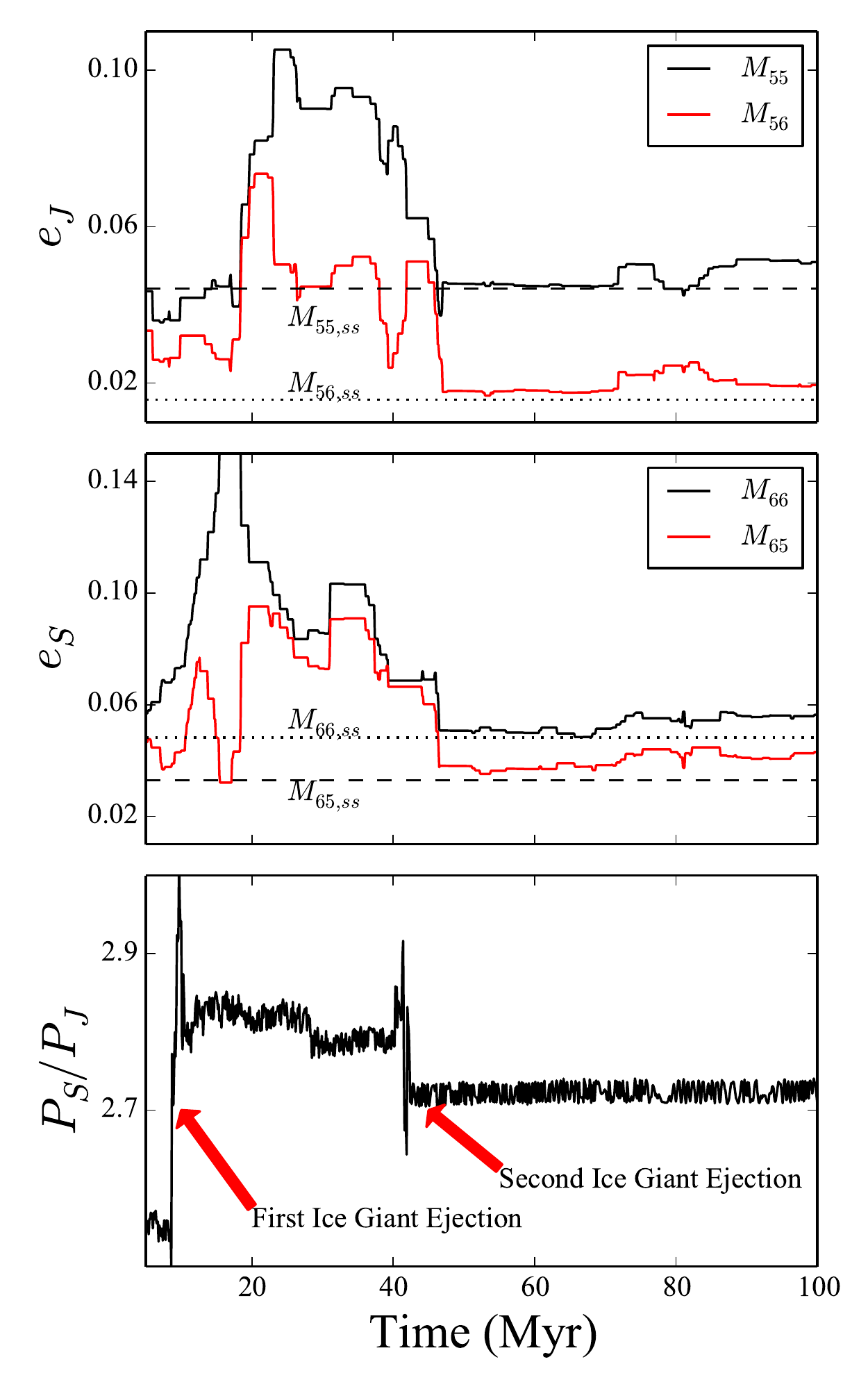}
	\includegraphics[width=.45\textwidth]{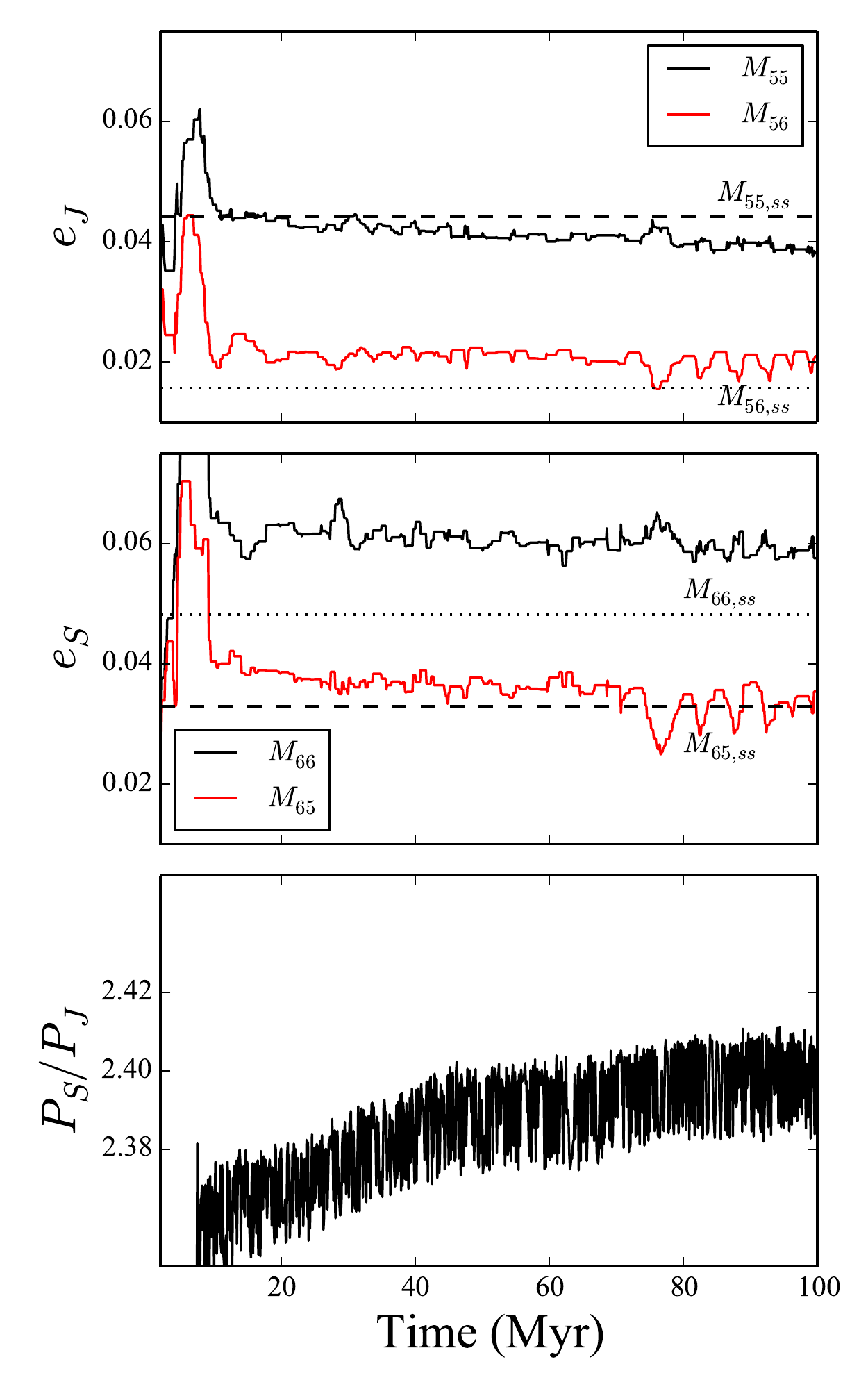}	
	\caption{Evolution of secular magnitudes ($M_{ij}$) of the Jupiter-Saturn system during and after the instability.  \textbf{Left panel:} An example simulation where a second, delayed instability and ice giant ejection further alters the planets' eccentric magnitudes, thereby yielding a satisfactory result for criterion \textbf{C}.  Note, however, that this particular system fails criterion \textbf{D} by exceeding $P_{S}/P_{J}=$2.5.  \textbf{Right Panel:} A more typical example of a criterion \textbf{C} satisfying simulation.  In this example all 4 secular magnitudes are excited to around twice their modern magnitude during the instability.  The temporarily eccentric orbits of Jupiter and Saturn rapidly damp within the first $\sim$10 Myr of the residual migration phase to close to the present-day values.  Conversely, damping in the amplitudes $M_{ij}$ final $\sim$80 Myr proceeds only at the $\sim$5-10$\%$ level.}
	\label{fig:modes}
\end{figure*}

Naively one might expect fully resolving residual migration to result in lower success rates for criterion \textbf{C} (the Jupiter-Saturn secular system) as the eccentric magnitudes tend to damp appreciably in this interval via dynamical friction \citep[see a more complete discussion in:][]{nesvorny12}.  Intriguingly though, our new \textit{100 Myr Low $M_{KB}$} simulations boast marginally improved rates of success for criterion \textbf{C} when compared to our reference case from Paper I (13$\%$ versus 10$\%$ of systems successful).  We investigated the cause of this discrepancy and determined that a small number of our new systems experienced an additional ice giant ejection and corresponding shake-up of the system's secular architecture after the 20 Myr point (i.e.: a five planet system at $t=$ 20 Myr transforms into a four planet system some time in the next 80 Myr).  An example of an evolution of this type is plotted in the left panel of figure \ref{fig:modes}.  This is figure is created by monitoring Jupiter and Saturn's maximum and minimum eccentricities throughout the simulation with a rolling 200 Kyr time window.  Thus, spurious fluctuations in the $M_{ij}$ values depicted in this figure are a consequence of our simulation output cadence being insufficient to accurately compute the secular magnitudes for each output time.  In spite of artificially shifting the inner ice giant's mean anomaly to force an instability, the planets continue to migrate for some time before the innermost ice giant scatters off of Jupiter and Saturn around $t=$ 6 Myr.  This dynamical exchange excites the Jupiter-Saturn system eigenmodes, $M_{ij}$ ($i,j=$ 5, 6), substantially above their current values, and similarly drives Saturn's semi-major axis beyond $P_{S}/P_{J}\simeq$ 2.8 \citep[note that this represents a poor solar system analog, and that such a final value of $P_{S}/P_{J}$ would have negative consequences for the asteroid belt:][]{deienno18,clement18_ab}.  Thus, at the 20 Myr point (our stop time in Paper I), this system is unsuccessful in terms of all four success criteria.  However, after an additional $\sim$25 Myr sequence of residual migration the third ice giant is ejected (thus satisfying criterion \textbf{A} as $N_{GP}=$ 4).  The removal of this additional ice giant's eccentric perturbations on the Jupiter-Saturn system calms their eccentricities, and the final system satisfies criterion \textbf{C} by yielding a remarkable analog of the modern planets' secular structure.  While this peculiar evolution is interesting from a dynamical standpoint, our simulation batch did not yield an example of a late ice giant ejection favorably altering the Jupiter-Saturn secular system that satisfied criterion \textbf{D} ($P_{S}/P_{J}$).  Thus, it seems unlikely that this represents a plausible evolutionary pathway for the solar system.  However, this result does not speak against the general viability of six planet configurations as the majority of such systems that finish with $N_{GP}=$ 4 eject both additional ice giants in rapid succession during the instability.

For comparison, an example of the evolution of the magnitudes, $M_{ij}$ ($i,j=$ 5, 6), in a system satisfying all four of our constraints is plotted in the right panel of figure \ref{fig:modes}.  While residual migration continues to appreciably alter $P_{S}/P_{J}$ for nearly the entire simulation duration (bottom panel), the majority of the post-instability damping in the amplitudes $M_{ij}$ occurs in the immediate few Myr following the instability.

\subsection{Improved planetesimal disk particle resolution}

Our simulations incorporating 5,000 planetesimals in the primordial Kuiper Belt (denoted \textit{High-res} in table \ref{table:results}) confirm the main findings of Paper I.  The primordial, eccentricity-pumped 2:1 Jupiter-Saturn resonance systematically improves the likelihood of replicating the two planets' modern configuration, while the final orbits of Uranus and Neptune are largely dependent on their initial orientation (i.e.: their masses and mutual resonances) and certain properties of the early planetesimal disk.

Our higher-resolution simulations finish with systematically worse success rates for criterion \textbf{A} when compared to our reference control runs from Paper I.  The source of this discrepancy is an increased fraction of $N_{GP}=$ 3 systems.  Indeed, 63 of 181 simulations finish with just one surviving ice giant, compared to just 36 in our \textit{100 Myr Low $M_{KB}$} set (which include instances of extremely late losses that are not possible in our $High-res$ batch by virtue of the shorter integration time).  Moreover, the evolution of our \textit{High-res} simulations largely bifurcate from that of the our other control runs after the instability.  Strikingly, nearly all (59 of 63) of the three planet systems in our \textit{High-res} case lose their final planet after $t=$ 10 Myr.  For comparison, less than half of our reference simulations from Paper I lose a planet after the 10 Myr point.  Without additional suites of simulations for comparison, it is unclear whether or not this result is a statistical artifact and a consequence of the instability's stochastic nature.  It is possible that the lower resolution simulations allow a larger random walk in phase space that tends to enable systems to avoid ejections that might have otherwise occurred.  While this result might be of interest for future investigations attempting to constrain the size frequency distribution of the primordial Kuiper Belt, we argue that it does not strongly speak against our proposed scenario.  In particular, the rates of success for criteria \textbf{C} and \textbf{D} (that comprehensively select for proper analogs of the Jupiter-Saturn system) are nearly identical in our \textit{Control}, \textit{100 Myr Low $M_{KB}$}, and \textit{High-res} runs.  Moreover, a reasonable fraction of systems (46$\%$) still finish with $N_{GP}=$ 4.

\subsection{High-eccentricity runs}

\begin{figure}
	\centering
	\includegraphics[width=.5\textwidth]{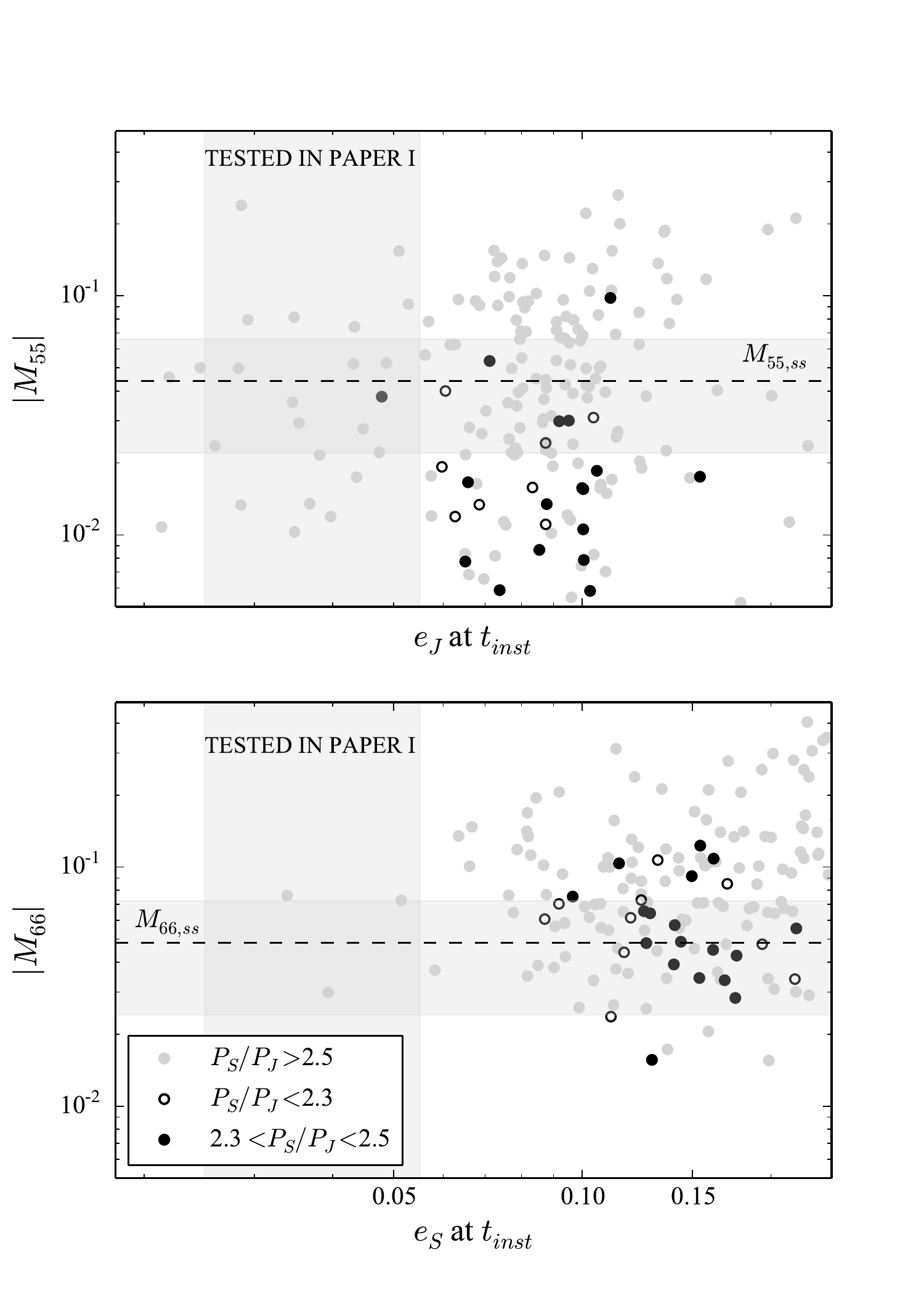}
	\caption{Jupiter and Saturn's eccentricities at the time of the instability in our various High-$e$ simulations (table \ref{table:ics}) versus their resultant eccentric magnitudes $M_{55}$ and $M_{66}$.  The shade of each point represents the simulations final $P_{S}/P_{J}$ value, with the most successful outcomes (2.3 $<P_{S}/P_{J}<$ 2.5) in black.  The vertical grey shaded region represents the parameter space of initial eccentricities for the planets probed in Paper I.  The dashed horizontal lines and parallel shaded regions represent the modern values of the planets' respective secular eccentric magnitudes, and the range of acceptable outcomes for criterion \textbf{C}, respectively.  The successful simulations are therefore those represented by black points falling in the horizontal shaded region.}
	\label{fig:tinst}	
\end{figure}

The success rates for our additional batch of high-eccentricity runs (table \ref{table:results}) are rather remarkable considering the fact that many of these instabilities ensue expeditiously since Jupiter and Saturn originate on near-crossing orbits; thus boosting the probability of the planets' entering the scattering regime \citep[e.g.:][]{ray09b,ray10}.  Moreover, our two sets of \textit{High-e} configurations produce nearly identical statistical results.  Thus, initializing the innermost ice giant on a crossing, resonant orbit with Saturn does not appreciably affect the systems' overall chances of success.  Figure \ref{fig:tinst} plots the distribution of final $M_{55}$ and $M_{66}$ values against Jupiter and Saturn's eccentricity at the time of the instability.  Realizations where the instability ensues from relatively low values of $e_{J}$ and $e_{S}$ are thus necessarily those where the planets' smoothly migrate for a significant period of time prior to the instability.  As this process tends to damp their orbits to near-zero eccentricity (we provide a more detailed discussion of this mechanism in Paper I), simulations that undergo significant pre-instability damping tend to correlate loosely with lower final $M_{ij}$ and $P_{S}/P_{J}$ values (see, for example, the depressed success rates for criterion \textbf{C} in our \textit{100 Myr High $M_{KB}$} set).  Conversely, instabilities that develop expeditiously almost exclusively yield violent evolutions, excessive Jupiter-Saturn period ratios, and extreme final eccentricities (grey points in the upper right corner of figure \ref{fig:tinst}).  

The damping of Jupiter's eccentricity prior to the instability towards the values examined in Paper I (0.025-05) is a key predictor of our simulations' success in terms of criterion \textbf{C} and \textbf{D} (top panel).  Contrarily, reasonable outcomes are achieved in systems' where the instability ensues at a point where Saturn's eccentricity is still relatively high (bottom panel).  In these simulations, Saturn's eccentricity damps rapidly after the event's onset via residual migration and, in some cases, interactions with the ejected ice giants that coincidentally occur at geometries favorable for de-exciting $e_{S}$. Thus, it is clear that a broad spectrum of potentially viable parameter space was left unexplored in Paper I (vertical grey shaded regions in figure \ref{fig:tinst}); particularly in terms of the range of initial values of $e_{S}$ that are feasible.  This result is particularly encouraging as it demonstrates the viability of our scenario in simulations originating from eccentricities similar to those found for 2:1 Jupiter-Saturn capture in \citet{pierens14} without relying on an artificial instability trigger.

\begin{figure*}
	\centering
	\includegraphics[width=.8\textwidth]{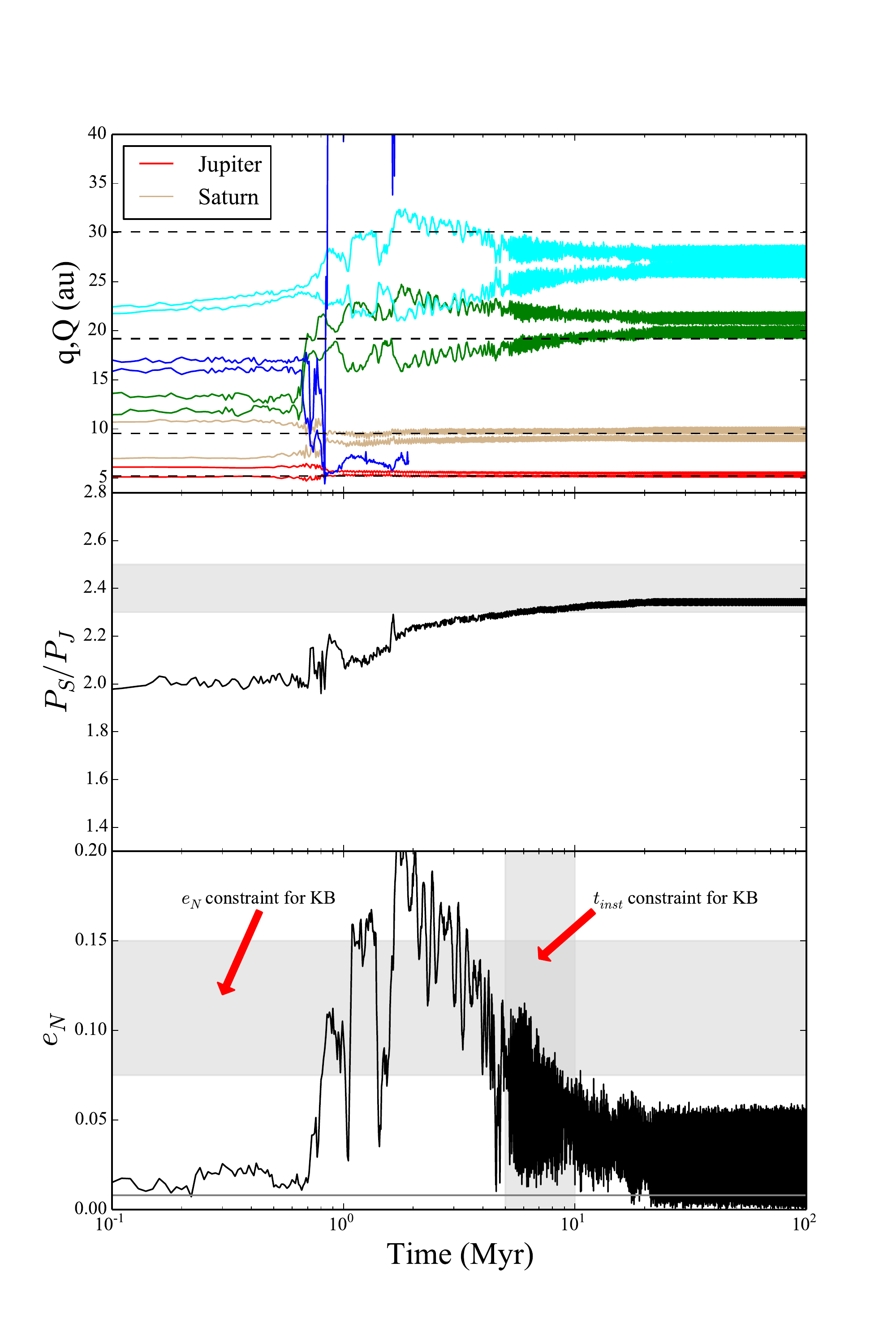}
	\caption{Example instability evolution beginning with five planets in a 2:1,2:1,3:2,3:2 resonant chain.  The simulation finished with $P_{S}/P_{J}=$ 2.35, $M_{55}=$.031, $M_{56}=$.015, $M_{65}=$.027, $M_{66}=$.043 (all four success criteria are satisfied).  The top panel plots the perihelion and aphelion of each planet over the length of the simulation.  The middle panel shows the Jupiter-Saturn period ratio.  The horizontal dashed lines in the upper panel indicate the locations of the giant planets' current semi-major axes.  The shaded region in the middle panel delimits the range of 2.3 $<P_{S}/P_{J}<$ 2.5.  The bottom panel plots the evolution of Neptune's eccentricity.  The horizontal shaded region denotes a range around $e_{N}\simeq$0.1; the preferred early migration eccentricity of \citet{nesvorny21_e_nep}, and the vertical shaded region delimits the corresponding preferred instability time (this should be interpreted as a minimum value; $\sim$5-10 Myr, as longer migration timescales are also compatible with Kuiper Belt constraints).  The horizontal grey line in the bottom panel represent's Neptune's modern eccentricity.}
	\label{fig:5gp_good}
\end{figure*}

A potential weakness of our high-eccentricity evolutions are the somewhat exotic resultant migration schemes of Neptune that are potentially inconsistent with certain constraints inferred from the observed Kuiper Belt.  Neptune's semi-major axis and eccentricity evolution both before and after the giant planet instability has been studied by several recent authors.  An important constraint on Neptune's early eccentricity evolution comes from the disparate eccentricity distributions of the hot and cold populations of KBOs \citep{dawson12,wolff12}.  In a similar manner, \citet{nesvorny15a} argued that Neptune's pre-instability migration was slow and non-eccentric ($\tau \gtrsim$ 10 Myr, $e_{N} \lesssim$ 0.1) in order to reconcile the inclination distribution of the hot Kuiper Belt.  Subsequent work in \citet{nesvorny15b} favored the migration of Neptune to as far as $\sim$28 au \textit{prior to} the instability, followed by a jump in semi-major axis to explain the so-called kernel of cold KBOs with $a\simeq$ 42-45 au.  More recently \citet{gomes18} and \citet{volk19} argued that particular combinations of $\tau$ and $e_{N}$ do not necessarily correlate with specific outcomes in terms of the replication of the Kuiper Belt's inclination distribution.  In particular, it might be possible that eccentric ($e_{N}\simeq$ 0.1) early ($\tau \lesssim$ 10 Myr) migration of Neptune was primarily responsible for sculpting the inclinations of hot KBOs \citep{nesvory20_i_kb,nesvorny21_e_nep}.  

Our proposed scenario axiomatically implies an eccentric early migration phase for Neptune as Jupiter and Saturn's dynamical excitation bleeds out to the ice giants rather precipitously in our eccentricity-pumped resonant chains.  However, the early migration values of $e_{N}$ ($\simeq$ 0.1-0.2) and instability times in our realizations are rather extreme compared to those proposed in past work.  Indeed, the median $\tau$ for our criteria \textbf{D} satisfying \textit{High-e} simulations is 1.2 Myr.  Figure \ref{fig:5gp_good} plots an example of a simulation from our \textit{5GP High-e} batch that simultaneously satisfies all four of our constraints.  Although the final value of $e_{Nep}$ in this simulation damps to near the solar system value, it is clear that Neptune's pre-instability migration is brief ($\sim$ 2 Myr) and its maximal eccentricity is rather high (in excess of $e_{N}\simeq$ 0.25 in this example).  Thus, future work must scrutinize whether such eccentric evolutions are compatible with key constraints from the Kuiper Belt's inclination distribution \citep{nesvorny15a}, resonant constituencies \citep{dawson12,kaib16,nesvorny16} and cold kernel population \citep{nesvorny15b,gomes18,gomes21}.

At this point it is worth briefly discussing the limits on the range of plausible initial eccentricities for Jupiter and Saturn.  In particular, we stress that our results should not be interpreted as evidencing a non-correlation between the planets' initial eccentricities and ultimate instability outcomes.  Indeed, the results of our current investigation coupled with those from Paper I demonstrate the importance or the planets' eccentricities damping considerably prior to the instability from the primordial values predicted in disk models of the 2:1 Jupiter-Saturn resonance capture \citep{pierens14}.  However,  some degree of primordial eccentricity-excitation is essential for the 2:1 resonance's viability.  In Paper I we studied a set of 2:1 instabilities where the gas giants initially inhabited circular orbits and found no final systems possessed adequately excited $M_{55}$ magnitudes and $P_{S}/P_{J}<$ 2.5.  On the opposite end of the eccentricity spectrum, we were unable to generate resonant chains with $e_{J}\gtrsim$ 0.10 and $e_{S}\gtrsim$ 0.30 that did not rapidly decompose and produce a violent Jupiter-Saturn scattering event \citep[e.g.:][]{ray09b,ray10}.  While we did not experiment with alternative methodologies for producing eccentric chains of resonant planets, we contend that this result evinces a firm upper limit on the range of feasible initial values of $e_{J}$ and $e_{S}$.  We also note that an additional primordial ice giant (or two) is essential for the success of our proposed scenario.  In Paper I we considered a moderately eccentric ($e_{J}=e_{S}=$ 0.025), 2:1,4:3,4:3 chain of four planets and found no final systems retained the appropriate number of outer planets.  During our current investigation we experimented with a less compact, \textit{High-e} ($e_{J}\simeq$ 0.10; $e_{S}\simeq$ 0.25) 2:1,2:1,2:1 chain and again found a null population of $N_{GP}=$ 4 systems.  Thus, the combined results from Paper I and our present manuscript lead us to constrain the range of viable parameter space for the primordial 2:1 Jupiter-Saturn resonance to five and six planet chains with $e_{J}\lesssim$ 0.10 and $e_{S}\lesssim$ 0.30 after nebular gas dispersal, and $e_{J} \simeq e_{S} \gtrsim$ 0.025 at the time of the instability. 

\section{Discussion and Conclusions}

In this paper we presented a supplementary batch of simulations investigating several open-ended questions from recent work reported in \citet{clement21_instb}.  Specifically, our work investigates an evolutionary scenario for the solar system where Jupiter and Saturn emerge from the nebular gas locked in a 2:1 MMR with inflated eccentricities.  While conventional models assuming a primordial 3:2 resonance between the gas giants struggle to adequately excite Jupiter's fifth eccentric mode ($M_{55}$) without over-exciting Saturn's forced eccentricity ($M_{56}$) and scattering Saturn into the distant solar system, our scenario provides a promising means of more consistently replicating the Jupiter-Saturn system.  In this manuscript, we scrutinized the \citet{clement21_instb} scenario with several batches of simulations incorporating longer integration timescales that fully resolve residual migration, more realistic primordial KBOs possessing masses similar to that of Pluto, and authentic initial conditions derived from hydrodynamical disk models in \citet{pierens14}.  

Our new results largely confirm the initial findings of \citet{clement21_instb}.  In particular, we conclude that the residual phase of migration (20-100 Myr after the instability) only minimally damps the eccentric magnitudes of the Jupiter-Saturn system.  While a sizable number of simulations experience additional migration that drives Saturn past its' present-day orientation with respect to Jupiter ($P_{S}/P_{J}=$ 2.49), the solar system remains well within the spectrum of outcomes generated in our simulations provided the initial total mass of the primordial Kuiper Belt is not excessively large.  Additionally, we note that the assumed particle resolution in the primordial external planetesimal disk does not qualitatively alter the statistical distribution of final Jupiter-Saturn configurations.  Lastly, we present an intriguing batch of simulations where Jupiter and Saturn begin on highly eccentric orbits ($e_{J}\simeq$ 0.10; $e_{S}\simeq$ 0.25) consistent with disk model studies of the planets' capture in the mutual 2:1 resonance \citep{pierens14} that do not utilize an artificial instability trigger.  Surprisingly, this batch of simulations produces many successful realizations.  In particular, when Jupiter's eccentricity damps slightly prior to the instability via dynamical friction, the overall results are effectively the same as those presented in \citet{clement21_instb}.  Future work should further validate the primordial 2:1 Jupiter-Saturn resonance by scrutinizing high-resolution simulations against constraints from the solar system's small body populations \citep[e.g.:][]{nesvorny15a,nesvorny15b,nesvorny16,izidoro16,deienno18,clement18_ab}.

\section*{Acknowledgments}

The authors thank Rodney Gomes and Ramon Brasser for insightful reviews of the manuscript.  R.D. acknowledges support from NASA Emerging Worlds program, grant $\#$80NSSC21K0387.  N.A.K. thanks the National Science Foundation for support under award AST-1615975 and NSF CAREER award 1846388.  A.I. acknowledges NASA grant 80NSSC18K0828 to Rajdeep Dasgupta, during preparation and submission of the work.  S.N.R. acknowledges support from the CNRS’s PNP program and NASA Astrobiology Institute's Virtual Planetary Laboratory Lead Team, funded via the NASA Astrobiology Institute under solicitation NNH12ZDA002C and cooperative agreement no. NNA13AA93A.  The majority of computing for this project was performed at the OU Supercomputing Center for Education and Research (OSCER) at the University of Oklahoma (OU).  The authors acknowledge the Texas Advanced Computing Center (TACC) at The University of Texas at Austin for providing {HPC, visualization, database, or grid} resources that have contributed to the research results reported within this paper. URL: http://www.tacc.utexas.edu.
\bibliographystyle{apj}
\newcommand{\sci}{$Science$ }
\bibliography{e55_2.bib}
\end{document}